\documentstyle[11pt]{article}
\begin{document}
\newcommand{\be}{\begin{equation}}
\newcommand{\ee}{\end{equation}}
\newcommand{\bea}{\begin{eqnarray}}
\newcommand{\eea}{\end{eqnarray}}
\newcommand{\beaa}{\begin{eqnarray*}}
\newcommand{\eeaa}{\end{eqnarray*}}
\newcommand{\qd}{\quad}
\newcommand{\qqd}{\qquad}
\newcommand{\npb}{\nopagebreak[1]}
\newcommand{\nn}{\nonumber}

\title{\bf Lattice and q-difference Darboux-Zakharov-Manakov
systems via $\bar{\partial}$-dressing method}
\author{L.V. Bogdanov\thanks{Permanent address:IINS,
Landau Institute for Theoretical Physics, Kosygin str. 2,
Moscow 117940, GSP1, Russia; e-mail Leonid@landau.ac.ru}
\hspace{0.1em} and B.G. Konopelchenko\\
Consortium EINSTEIN\thanks{European Institute for
Nonlinear Studies via Transnationally Extended Interchanges},
\\ Dipartimento di Fisica dell'Universit\`a
and Sezione INFN, \\73100 Lecce, Italy
}
\date{}
\maketitle
\begin{abstract}
A general scheme is proposed for introduction of lattice and
q-difference variables to integrable hierarchies in frame
of $\bar{\partial}$-dressing method .
Using this scheme, lattice and q-difference Darboux-Zakharov-Manakov
systems of equations are derived. Darboux, B\"acklund
and Combescure transformations and exact
solutions for these systems are studied.
\end{abstract}
\section{Introduction}In the present paper we will discuss the lattice
and q-difference
integrable versions of the well-known Darboux-Zakharov-Manakov
(DZM) system
\be
\partial_i\partial_j H_k=
(\partial_j H_i)H_i^{-1}
\partial_i H_k
+(\partial_i H_j)H_j^{-1} \partial_j H_k
\label{Darboux}
\ee
where $i,j,k=1,2,3$ and $i\neq j\neq k\neq i$.
The system (\ref{Darboux}) is about one hundred years old.
It has been discovered by G. Darboux within his study of the
triply conjugate systems of surfaces \cite{Darboux} and then has been
studied intensively by geometrists (see e.g. \cite{GEOM}-\cite{GEOMB}).

The system (\ref{Darboux}) has been rediscovered 10 years ago
by Zakharov and Manakov within the framework of dressing method
\cite{ZM1}. But only few years ago the interrelation has been revealed
between the new results of Zakharov and Manakov and the old
geometrical constructions of Darboux \cite{Dryuma}.
During the last years the DZM system has been studied in detail:
wide classes of exact solutions have been constructed \cite{Kon1},
different transformations and reductions have been analyzed
\cite{Kon2}, \cite{Schief}.

It is remarkable that the applications of the DZM system are not exhausted
only by the differential geometry. It was discovered recently
that the system (\ref{Darboux}) plays a key role in the theory of the
Hamiltonian and semi-Hamiltonian systems of hydrodynamical type
\cite{Tsarev}-\cite{Tsarev1} and in
two-dimensional topological field theories
\cite{Dubrovin}. The DZM system (\ref{Darboux}) arises also as
the universal equations for the certain hierarchies of integrable
equations \cite{Nijhoff}, \cite{BM}.

Our study of the difference and q-difference versions of the
DZM system is motivated by the well established understanding that the
difference versions of integrable systems reveal the deeper nature and
algebraic structure of the corresponding continuous
nonlinear integrable PDEs.

In the present paper we construct the difference and q-difference DZM
systems using the $\bar{\partial}$-dressing method \cite{ZM1},
\cite{BM}, \cite{ZM1}
(see also \cite{Kon1}).
We discuss the first order form of the difference
and q-difference DZM system and its properties. Darboux, B\"acklund
and Combescure  transformations for the DZM system are derived
via $\bar{\partial}$-dressing. Exact solution are found.

We derive also the analogue of the Hirota bilinear identity for the
difference and q-difference DZM systems (see \cite{NLS}) via
the $\bar{\partial}$-dressing method. Such bilinear identities can be
used as the starting point of the Hirota-Sato approach to the
DZM hierarchies.

\section{Lattice and q-difference variables
in $\bar{\partial}$-dressing formalism}

The scheme of the $\bar{\partial}$-dressing method uses the nonlocal
$\bar{\partial}$-problem with the special dependence
of the kernel on additional
variables
\begin{eqnarray}\
\bar{\partial}(\chi({\bf x} ,\lambda)-\eta({\bf x},\lambda))=
\int\!\!\!\int_{\bf C}\: d\mu\wedge d\bar{\mu}\chi(\mu)g^{-1}(\mu)
R(\mu,\lambda)g(\lambda))
, \label{dpr} \\
(\chi({\bf x} ,\lambda)-\eta({\bf x},\lambda))_
{|\lambda|\rightarrow\infty}\rightarrow 0.\nn
\end{eqnarray}
where $\lambda \in {\bf C},$  $ \bar{\partial}={\partial /
\partial \bar{\lambda}}$,
$\eta({\bf x},\lambda)$ is a rational function
of $\lambda$ (normalization). In this work we treat
non-commutative case, so the function $\chi(\lambda)$
and the kernel $R(\lambda,\mu)$ are matrix-valued
functions.

A dependence of the solution
$\chi(\lambda)$ of the problem (\ref{dpr}) on extra variables
is hidden in the function $g(\lambda)$.
Usually these variables are continuous space and
time variables, but it is possible also to introduce discrete (lattice)
an q-difference variables into $\bar{\partial}$-dressing formalism.
We will consider the
following functions $g(\lambda)$
\begin{eqnarray}
g^{-1}_i&=&\exp(K_i x_i)\label{c}
;\quad
{\partial\over\partial x_i}g^{-1}=K_i g^{-1},\\
g^{-1}_i&=&(1+l_i K_i)^{n_i}\label{d};\:
\Delta_i g^{-1}={g^{-1}(n_i+1)-g^{-1}(n_i)\over
l_i}=K_i g^{-1},
\\
g^{-1}_i&=&{\rm e}_q (K_i y_i);\label{q}\quad
\delta_i^q g^{-1}={g^{-1}(qy_i)-g^{-1}(y_i)\over
(q-1)y_i}=K_i g^{-1}.
\end{eqnarray}
Here $K_i(\lambda)$ are meromorphic matrix functions
commuting for different values of $i$. The function
(\ref{c}) introduces a dependence on continuous variable
$x_i$, the function (\ref{d}) -- on discrete variable $n_i$
and the function (\ref{q}) defines a dependence of $\chi(\lambda)$
on the variable $y_i$ (we will call it a q-difference variable).
To introduce a dependence on several variables (may be of different
type), one should consider a product of corresponding functions $g(\lambda)$
(all of them commute). Equations in the right part of
(\ref{c}-\ref{q}) and the boundary condition $g(0)=1$
characterize the corresponding functions (and give a definition
of ${\rm e_q(y)}$).
These equations
play a crucial role in the algebraic scheme of constructing
integrable equations in frame of $\bar{\partial}$ dressing method.
This scheme is based on the assumption of unique solvability
of the problem (\ref{dpr}) and on the existence of special
operators, which transform solutions
of the problem (\ref{dpr}) into the solutions of the same problem
with other normalization.

We suppose that the kernel $R(\lambda,\mu)$
equals to zero in some open subset $G$  of
the complex plane
with respect to $\lambda$ and to $\mu$.
This subset should typically include all zeroes
and poles of the considered class of functions
$g(\lambda)$ and a neighborhood of infinity.

In this case the solution of the problem (1)
normalized by $\eta$
is the function \[\chi(\lambda)=
\eta({\bf x},\lambda)+\varphi({\bf x},\lambda),\]
where $\eta(\lambda)$ is a rational function
of $\lambda$ (normalization), all poles of $\eta(\lambda)$
belong to $G$,
$\varphi(\lambda)$
decreases as $\lambda \rightarrow \infty$ and is
{\em analytic} in $G$.

The solutions of the problem (\ref{dpr}) with a
rational normalization form a linear space,
let us denote this space $W$. This space depends
on corresponding extra variables (in fact it if a functional
of the function $g$).
It is easy to check that
\be
W(g)=gW(1)
\label{group}
\ee
The $\bar{\partial}$-problem (\ref{dpr}) implies the
difference and q-difference extensions of the famous
Hirota bilinear identity. Indeed, let us consider the
problem (\ref{dpr}) and its formally adjoint for
the function normalized by $(\lambda-\mu)^{-1}$
with different functions $g$ (i.e. with different values
of coordinates)
\begin{eqnarray}
{\partial\over\partial\bar{\lambda}}\chi(\lambda,\mu)
=2\pi i \delta(\lambda-\mu)+
\int\!\!\!\int_{\bf C} d\nu\wedge d\bar{\nu}\chi(\nu,\mu)g_1(\nu)^{-1}
R(\nu,\lambda)g_1(\lambda),\nn\\
{\partial\over\partial\bar{\lambda}}\chi^{\ast}(\lambda,\mu)
=-2\pi i \delta(\lambda-\mu)-
\int\!\!\!\int_{\bf C} d\nu\wedge d\bar{\nu}g_2(\mu)^{-1} R(\lambda,\nu)
g_2(\nu)\chi^{\ast}(\nu,\mu).
\end{eqnarray}
After simple calculations (in the case of continuous
variables see \cite{Carroll})
we obtain
\be
\int_{\gamma} \chi(\nu,\lambda;g_1)g_1^{-1}(\nu)g_2(\nu)
\chi^{\ast}(\nu,\mu;g_2)d\nu=0 ,
\label{HIROTA}
\ee
where $\gamma$ is the boundary of $G$. It follows from
(\ref{HIROTA}) that in $\bar{G}$ the function $\chi(\lambda,\mu)$
is equal to $\chi^{\ast}(\mu,\lambda)$, so in fact this
identity should be written for one function.  It
is possible to take identity (\ref{HIROTA})
instead of (\ref{dpr}) as a starting point for the
algebraic scheme of constructing equations.

\section{Derivation of DZM equations}
The algebraic scheme of constructing equations
is based on the following property
of the problem (\ref{dpr}) with the dressing functions
(\ref{c}-\ref{q}): if $\chi({\bf x},{\bf n},
{\bf y},\lambda)\in W({\bf x},{\bf n},
{\bf y})$,
then the functions
\bea
D_{i}^c \chi&=&\partial / \partial x_{i}\chi +\chi K_i(\lambda) \nn\\
D_{i}^d \chi&=&\Delta_{i}\chi +T_i\chi K_i(\lambda) \nn\\
D_{i}^q \chi&=&\delta^q_{i}\chi +T_i^q \chi K_i(\lambda)
\label{D}
\eea
also belong to $W$, where $Tf(n)=f(n+1)$,
$T^q f(y)=f(qy)$. We can multiply the
solution from the left
by the arbitrary matrix function
of additional variables,
$u({\bf x},{\bf n},{\bf y})\chi
\in W$. So the operators (\ref{D}) are the generators
of Zakharov-Manakov ring of operators, that transform
$W$ into itself.

Combining this property
with the unique solvability of the problem
(1), one obtains the differential relations
between the coefficients of expansion of functions
$\chi({\bf x},{\bf n}, {\bf y},\lambda)$ into powers of
$(\lambda-\lambda_{p})$ at the poles of
$K_{i}(\lambda)$ \cite{BM}.

The derivation of equations in this case is completely
analogous to the continuous case \cite{ZM1}.
First we choose three functions $K_i(\lambda)$, $K_j(\lambda)$,
$K_k(\lambda)$ in the form
\be
K_i(\lambda)={A_i\over \lambda - \lambda_i}
\ee
where $A_i$, $A_j$, $A_k$ are commuting matrices,
$\lambda_i\neq\lambda_j\neq\lambda_k\neq\lambda_i$.
Then we introduce the solution of the problem (\ref{dpr})
$\chi(\lambda)$ with the canonical normalization ($\eta(\lambda)=1$).
The following derivation will be conducted for q-difference case,
to get the difference case you should just change
$\delta^q_i$ for $\Delta_i$ and $T_i$ for $T^q_i$.

The function $\chi$ satisfies the linear equations
\be
D_i^q D_j^q\chi=
T_i^q((D_j^q(\lambda_i)\chi_i)\chi_i^{-1})
D_i^q (\lambda_k)\chi\nn
\\+T_j^q((D_i^q(\lambda_j)\chi_j)\chi_j^{-1}) D_j^q (\lambda_k)\chi
\label{ZM}
\ee
where $\chi_i=\chi(\lambda_i)$. Evaluating the equation
(\ref{ZM}) at the point $\lambda_k$, we obtain the closed system of
equations for the functions $\chi_i$
\be
D_i^q(\lambda_k) D_j^q(\lambda_k)\chi_k=
T_i^q((D_j^q(\lambda_i)\chi_i)\chi_i^{-1})
D_i^q (\lambda_k)\chi_k
+T_j^q((D_i^q(\lambda_j)\chi_j)\chi_j^{-1}) D_j^q (\lambda_k)\chi_k.
\label{ZME}
\ee
It is possible to transform the operators $D$ to
usual derivatives or difference operators by the substitution
\be
\chi_k=H_k g_i(\lambda_k) g_j(\lambda_k); \quad
\psi=\chi g_i(\lambda)g_j(\lambda)g_k(\lambda)
\label{substitution}
\ee
(this substitution works for all cases - continuous, difference and
q-difference, you should only take the corresponding function
(\ref{c}-\ref{q})). Then the linear equations
(\ref{ZM}) and the equations for the functions
$H_i$ read
\bea
\delta_i^q\delta_j^q\psi&=&
T_i^q((\delta_j^qH_i){H_i}^{-1})
\delta_i^q \psi
+T_j^q((\delta_i^qH_j){H_j}^{-1}) \delta_j^q
\psi,
\label{ZMEL}\\
\delta_i^q\delta_j^qH_k&=&
T_i^q((\delta_j^qH_i){H_i}^{-1})
\delta_i^q H_k
+T_j^q((\delta_i^qH_j){H_j}^{-1}) \delta_j^q
H_k.
\label{ZME1}
\eea
The equations  (\ref{ZME1}) represent a
q-difference integrable deformation of
Zakharov-Manakov system.

The system (\ref{ZME1}) can be rewritten in the
first order form by the substitution
\be
\beta_{ij}=(T^q_i H_i)^{-1}\delta^q_i H_j.
\ee
Using the identity
$
\delta^q(g^{-1})=-(T^q g)^{-1}(\delta^{q}g)g^{-1}
$
we obtain the equations
\be
\delta^q_k \beta_{ij}=(T^q_k \beta_{ik})\beta_{kj}.
\label{ZME2}
\ee
Correspondingly the linear system (\ref{ZMEL}) becomes
\be
\delta^q_k \psi_i=(T^q_k \beta_{ik})\psi_k
\label{ZME3}
\ee
where $\psi_i=(T^q_i H_i)^{-1}\delta^q_i\psi$.
Note that the equations (\ref{ZME2}) and (\ref{ZME3})
can be derived directly from the bilinear identity (\ref{HIROTA}).
In this case the subset $G$ consists of the neighborhoods of
the points $\lambda_i,\lambda_j,\lambda_k$ and $\lambda=\infty$.

The continuous version of the system (\ref{ZME2}) has important
applications in the theory of systems of hydrodynamical type
and topological field theory \cite{Dubrovin}. We hope that
similar applications will be found for equations (\ref{ZME2}) too.

\section{Darboux transformation}
Now we will demonstrate how certain symmetry transformations for the
DZM system can be derived via $\bar{\partial}$-dressing method.

Let us introduce the function $\widetilde{g}(\lambda)$ in addition
to the functions $g_i, g_j, g_k$
\be
\widetilde g(\lambda)={\lambda - \lambda_i \over \lambda -
\widetilde{\lambda}}
\label{Bgroup}
\ee
where $\widetilde{\lambda}\in G$. It follows from (\ref{group})
that
\be
\widetilde{W}(y_i,y_j,y_k)=\widetilde{g}W(y_i,y_j,y_k)
\label{group1}
\ee
Let the canonically normalized function $\chi$ in the space $W$
be given; it satisfies linear equations (\ref{ZM}). The
property (\ref{group1}) gives an opportunity to calculate
canonically normalized function $\widetilde{\chi}$ in the space
$\widetilde{W}$ in terms of $\chi$,
this function also satisfies equations (\ref{ZM})
(with other potentials).
It implies that
\be
\widetilde{\chi}= u({\bf y}) \widetilde{g} \chi + v({\bf y}) \widetilde{g}
D^q_i\chi
\ee
with the properly chosen functions $u$ and $v$.
Using two conditions: the absence of poles and and unit
asymptotics at infinity, we get
\be
\widetilde{\chi}=   \widetilde{g}(\chi-
\chi(\widetilde{\lambda})(D^q_i\chi)^{-1}(\widetilde{\lambda})D^q_i\chi)
\ee
If one transforms operators $D^q_i$ into q-difference operators
$\delta^q_i$ by the substitution
$
\chi=\psi g_i g_j g_k,
$
one obtains
\be
\widetilde{\psi}(\lambda)=\widetilde{g}(1-
\psi(\widetilde{\lambda})(\delta^q_i\psi)^{-1}(\widetilde{\lambda})\delta^q_i)
\psi(\lambda)
\ee
that is the well-known Darboux transformation for the DZM
system (see \cite{Darboux}, \cite{Kon2}, \cite{Schief})

\section{B\"acklund transformation}

Introduction of group element (\ref{Bgroup}) may be treated
in a different manner, namely, as introduction of extra discrete variable
with
\be
\widetilde D=\widetilde \Delta +\widetilde K \widetilde T;\quad
\widetilde K= {\lambda_i-\widetilde{\lambda}\over \lambda-\lambda_i}
\ee
In these notations $\widetilde \chi=\widetilde T \chi$.
Using the formulae (\ref{ZME1}) for two q-difference
variables $y_i$, $y_j$ and discrete variable $\widetilde n$,
we obtain the following transformation
for the q-deformation of DZM system (\ref{ZME1})
\bea
\widetilde\Delta\delta_j^q H_i&=&
\widetilde T((\delta_j^q H_k){H_k}^{-1})
\widetilde\Delta H_i
+T_j^q((\widetilde\Delta H_j){H_j}^{-1}) \delta_j^q
H_i\\
\delta_i^q\widetilde\Delta H_j&=&
T_i^q((\widetilde\Delta H_i){H_i}^{-1})
\delta_i^q H_j
+\widetilde T((\delta_i^q H_k){H_k}^{-1}) \widetilde\Delta
H_j
\eea
This transformation establishes a connection between two
solutions of the system (\ref{ZME1}):
$H_i$, $H_j$, $H_k\rightarrow
\widetilde TH_i$, $\widetilde TH_j$, $\widetilde TH_k$ and
it is nothing but the B\"acklund transformation.

\section{Combescure transformation}

To derive Combescure transformation in frame of
$\bar{\partial}$-dressing method, it is
necessary to use freedom to choose a normalization
of the problem (\ref{dpr}) in quite a nontrivial way.
In this section we consider the commutative case
of $\bar{\partial}$-dressing method, so all
functions take their values in {\bf C}.
Let us introduce solution of the problem ($\ref{dpr}$) $\chi(\lambda,\mu)$
normalized by $(\lambda-\mu)^{-1}$,
where $\mu$ is a parameter, $\mu\in G$,
and let us modify operators $D$ just adding
constants $c_i={A_i\over(\lambda_i-\mu)}$ to them
\be
{D'}^q_i=D^q_i+c_i=D^q_i+{A_i\over (\lambda_i-\mu)}=
\delta^q_i-{A_i\over(\lambda_i-\mu)}(T^q_i-1) +
A_i{\lambda-\mu\over(\lambda-\lambda_i)(\lambda_i-\mu)}T^q_i
\ee
We would like to emphasize the kernel
of the problem (\ref{dpr}) remains the same. Then the function
$\chi(\lambda,\mu)$ satisfies the equation (\ref{ZME})
with the modified operators $D'$. To transform operators $D'$
to q-difference operators in this case one should use a substitution
\bea
\chi_k&=&H_k g_i(\lambda_k){\rm e}^{-1}_q(c_iy_i) g_j(\lambda_k){\rm
e}^{-1}_q(c_jy_j),\nn\\
\psi&=&\chi g_i(\lambda)g_j(\lambda)g_k(\lambda)
{\rm e}^{-1}_q(c_iy_i){\rm e}^{-1}_q(c_jy_j){\rm e}^{-1}_q(c_ky_k).
\label{substitution1}
\eea
So using different
normalizations we obtain different solutions of the system
(\ref{ZME1}). It happens that the connection between
these solutions is given
by the Combescure transformation. Indeed, unique
solvability of the problem (\ref{dpr}) implies
that
$
{D'}^q_i\chi(\lambda,\mu)=u_i({\bf y}) D^q_i\chi(\lambda)
$
or, after substitution (\ref{substitution1})
\be
{\delta}^q_i\psi(\lambda,\mu) =U_i({\bf y})\delta^q_i\psi(\lambda).
\label{COMB}
\ee
The compatibility conditions for equations (\ref{COMB}) give
the q-deformation of equations of Combescure transformation
\cite{Darboux}
\be
\delta^q_j U_i=(T_iU_j-T_jU_i)T_i^q((\delta_j^qH_i){H_i}^{-1}).
\ee
These equations in the continuous case are important
for the connection with the systems of hydrodynamical type
\cite{Tsarev}-\cite{Tsarev1}.

\section{Exact solutions}

The procedure of getting exact solutions in frame
of $\bar{\partial}$-dressing method is based on the fact
that for degenerate kernel $R(\lambda,\mu)$ the problem
(\ref{dpr}) is explicitly solvable. This property keeps
for the case of lattice and q-difference variables too.
Let us treat a simple example.

There is one important special case of
nonlocal $\bar{\partial}$-problem which is exactly solvable,
which corresponds to plane soliton solutions.
This is a case of $\delta$-functional kernels
\begin{equation}
R(\lambda,\mu)=2\pi i\sum_{\alpha=1}^N R_{\alpha}\delta
(\lambda-\lambda_{\alpha})
\delta (\mu-\mu_{\alpha}),
\end{equation}
where $\lambda_{\alpha}$, $\mu_{\alpha}$ is a set of points in the complex
plane,
$\lambda_{\alpha}\neq \mu_{\alpha'}$,
$$R_{\alpha}=(g_ig_jg_k)^{-1}(\mu_{\alpha})C_
{\alpha}g_ig_jg_k(\lambda_{\alpha}).$$
In this case the solution of the problem
(1) is a rational function, and the problem (1) reduces to the system
of linear equations. As a result the canonically normalized function
$\chi$ is given by
\begin{eqnarray}
\chi(\lambda)=1-
\sum_{\alpha,\alpha'}((A')^{-1})_{\alpha\alpha'}{1\over
(\lambda-\lambda_{\alpha})},
\label{rational2}\\
A'_{\alpha\alpha'}=R_{\alpha}^{-1}\delta _{\alpha\alpha'}
-{1\over \mu_{\alpha} -\lambda_{\alpha'}},\nn
\end{eqnarray}
To get solutions of the DZM equations (\ref{ZME1}) from
this formula, one should use relation $\chi_i=\chi(\lambda_i)$,
substitution (\ref{substitution}) and corresponding explicit
expressions for the functions $g_i$.
\subsection*{Acknowledgments}
This work was supported in part by INTAS (International
Association for the promotion of cooperation with scientists
from independent states of the former Soviet Union).
The first author (L. B.) also acknowledges partial
support by Soros Foundation (ISF) (grant MLY000)
and Russian Foundation for Fundamental Studies
(grant 94-01-00899).

\end{document}